# Spin and Dipole Ordering in $Ni_2InSbO_6$ and $Ni_2ScSbO_6$ with corundum-related structure.

**S.A. Ivanov** [a,b], **R. Mathieu** [b], **P. Nordblad** [b], **R. Tellgren**[c]**, C. Ritter** [d]**, E. Politova** [a], **G. Kaleva** [a], **A. Mosunov** [a]**, S. Stefanovich** [a]**, M. Weil** [e]

a- Center of Materials Science, Department of Inorganic Materials, Karpov' Institute of Physical Chemistry, Moscow, 105064, Russia

b- Department of Engineering Sciences, Uppsala University, Box 534, SE-751 21 Uppsala, Sweden

c- Department of Materials Chemistry, Uppsala University, Box 538, SE-751 21 Uppsala, Sweden

d -Institute Laue-Langevin, Grenoble, France

e- Institute for Chemical Technologies and Analytics, Vienna University of Technology, A-1060 Vienna, Austria

## Abstract

The complex metal oxides $Ni_2InSbO_6$ (NISO) and $Ni_2ScSbO_6$ (NSSO) have been prepared in form of polycrystalline powders by a solid state reaction route. The crystal structure and magnetic properties of the compounds were investigated using a combination of X-ray and neutron powder diffraction, electron microscopy, calorimetric and magnetic measurements. The compounds adopt a trigonal structure, space group *R3*, of the corundum related $Ni_3TeO_6$ (NTO) type. Only one of the octahedral Ni positions (Ni(2)) of the NTO structure was found to be occupied by In (Sc). NTO has non-centrosymmetric structure and is ferroelectric below 1000 K, dielectric and second harmonic measurements suggest that also NISO and NSSO are correspondingly ferroelectric. Magnetization measurements signified antiferromagnetic ordering below $T_N$=60 K (NSSO) and 76 K (NISO). The magnetic structure is formed by two antiferromagnetically coupled incommensurate helices with the spiral axis along the b-axis and propagation vector k = [0, $k_y$,0] with $k_y$= 0.036(1) (NSSO) and $k_y$= 0.029(1) (NISO). The observed structural and magnetic properties of NISO and NSSO are discussed and compared with those of NTO.

*Keywords:* Neutron scattering; Crystal structure; Magnetic properties; Ferroelectricity.

# 1. Introduction

The origin and understanding of the coupling phenomena between different physical properties within a material is a central subject of solid state science. High expectations on new physics and applications surround theoretical and experimental investigations of multiferroic compounds - materials possessing more than one order parameter, which can be of magnetic, electric and/or elastic nature [1]. The possible coexistence of and coupling between ferromagnetism and ferroelectricity in matter has been intensely investigated in recent years [2-9]. Existing multiferroic compounds belong to different crystallographic classes, and although some general rules governing their behaviour have been established [10,11], there is as yet no coherent understanding of multiferroic behaviour.

With a few exceptions, the multiferroic materials that have been discovered so far are transition metal oxides, mainly with a perovskite-related structure [12-15]. It has been found that some single-phase multiferroic materials possessing geometrically frustrated spin networks show a large magneto-electric effect [7-8]. A key factor in the ferroelectricity of these materials lies in their non-collinear spiral magnetic structures with a cycloidal component in which the magnetic structure itself breaks the inversion symmetry [9-10]. Understanding of these magnetic structures requires the use of the Goodenough–Kanamori–Anderson rules determining the sign and magnitude of exchange in insulators [16-18].

The complex metal oxides $A_3TeO_6$ (ATO, A=Mn, Co, Ni and Cu) have garnered interest due to their low temperature magnetic properties [19-23]. In the structures of ATO tellurates, the presence of crystallographically non-equivalent sites for magnetic cations provides extra degrees of freedom to manipulate the properties [24-27]. $Ni_3TeO_6$ (NTO) was first prepared as a small crystals and found isotypical with the corundum-related $Mg_3TeO_6$ family [28]. Later, Kosse et al. [29,30] identified several $A_2BSbO_6$ (A=Mg,Cd; B=Sc,In) oxides with same type of structure as possible ferroelectrics. However, many experimental and theoretical questions about this structural family still remain open [31-33]. Comprehensive studies of structural and magnetic properties of NTO in powder and single crystalline form have been previously been reported [34,35]. Peculiar to NTO is the lack of centre of symmetry and the presence of hexagonal planes with Ni spins that order at low temperature, both of which phenomena could induce ferroelectricity. NTO has been found to order antiferromagnetically below 52 K through ferromagnetically coupled planes alternating along the c axis [21-23] not allowing magnetically induced electric polarization.

However, Ni site substitution in the NTO structure emerges interesting, since the magnetism is dictated by the interactions within the Ni-O-Ni network; any substitution into this network may drastically modify the physical properties. As far as nonmagnetic impurities are concerned, one expects impurities to dilute the Ni-O-Ni network without introducing any additional magnetic interaction. However, apart from diluting the Ni-O-Ni network, such substitutions with appreciable ionic size mismatch may lead to changes of the magnetic interaction and structure.

In this communication, substitution of $Sc^{3+}$ ($In^{3+}$) - $Sb^{5+}$ for $Ni^{2+}$ - $Te^{6+}$ cations in NTO is attempted and ceramic samples of $Ni_2InSbO_6$ (NISO) and $Ni_2ScSbO_6$ (NSSO) are synthesized. The ionic radius of 6-coordinated $Sc^{3+}$ (0.745 A) and $In^{3+}$ (0.80 A) are larger than the radii of $Ni^{2+}$ (0.69 A), $Te^{6+}$ (0.56 A) and $Sb^{5+}$ (0.60 A). Thus, it is expected that Sb will reside in the Te sites because the ionic radius, oxidation state, and bonding preferences of $Te^{+6}$ are very similar to those of $Sb^{+5}$ [24-27] and that Sc and In will occupy Ni sites of the NTO structure. The structural, dielectric and magnetic properties of the samples are experimentally investigated and it is found that the compounds are ferroelectric below 1000 K and exhibit long-range incommensurate antiferromagnetic order at low temperatures.

## 2. Experimental

### 2.1 Sample preparation

#### 2.1.1 Single crystals

Single crystals of NTO were grown by chemical vapour transport reactions. X-ray amorphous $TeO_3$ was prepared by heating $H_6TeO_6$ (Merck, p.A.) at 400 °C for 12 hours; $PtCl_2$ was prepared according to reference [33]. A thoroughly ground mixture of NiO (Merck, p.A.) and $TeO_3$ in the stoichiometric ratio 3:1 was placed in a silica ampoule which was evacuated, sealed, and heated within 3 h to 973 K and kept at this temperature for one day. X-ray powder diffraction of the green microcrystalline material revealed a single phase product. 300 mg of this material was then mixed with 15 mg $PtCl_2$ and loaded in an evacuated and sealed silica ampoule (length 12 cm, diameter 1.2 cm) which was heated in a temperature gradient 1103 → 1023 K. $PtCl_2$ decomposes at that temperature with release of $Cl_2$ which then serves as the actual transport agent. After ten days, translucent green hexagonal plates with edge-lengths up to 1.5 mm had formed in the colder part of the ampoule.

#### 2.1.2 Ceramic samples

High quality ceramic samples of NSSO and NISO were prepared by solid state reaction. High purity NiO, $Sc_2O_3$, $In_2O_3$ and $Sb_2O_5$ were used as starting materials. The binary oxides were weighed in appropriate proportions for the requested formula using a high precision electronic balance. The homogenized stoichiometric mixtures were thoroughly mixed and ball-milled for several hours and then pressed in the form of cylindrical pellets with diameter 11 mm and thickness 5 mm. The sample preparation process included several stages of calcination, which efficiently suppressed the formation of undesirable stable impurity phases. The disks were first sintered at 700 C for 8 hours, grounded into fine powders, pressed and annealed again several times with temperature intervals of 100 C up to 1200 C with intermediate milling. The series of grinding and sintering procedures were performed until the XRD pattern showed the expected spectra without impurity lines. After the last ball-milling the final sintering was made at 1300° C for 24 h. The mixture was weighed before and after heat treatment to determine possible loss due to evaporation. In all cases, the weight difference was negligible (<0.01%). Possible parasitic phases can be leached in 10% diluted $HNO_3$.

## 2.2 Chemical composition

The chemical composition and the homogeneity of the prepared crystals and ceramic samples were analyzed by energy-dispersive spectroscopy (EDS) using a JEOL 840A scanning electron microscope and INCA 4.07 (Oxford Instruments) software. Analyses of different grains showed that the concentration ratios of Ni, Sc (In) and Sb are the stoichiometric ones within the instrumental resolution (0.05). The cation content of the samples prepared for diffraction measurements were determined by inductively coupled plasma atomic emission spectroscopy performed with an ARL Fisions 3410 spectrometer.

## 2.3 X-ray diffraction

Structural investigation of NTO single crystals was performed on a SMART Bruker diffractometer with results that are in a good agreement with earlier published data [35]. The phase identification and purity of the powder sample was checked from X-ray powder diffraction (XRPD) patterns obtained with a D-5000 diffractometer using Cu $K_\alpha$ radiation. The ceramic samples of NSSO and NISO were crushed into powder in an agate mortar and suspended in ethanol. A Si substrate was covered with several drops of the resulting suspension, leaving randomly oriented crystallites after drying. The XRPD data for Rietveld analysis were collected at room temperature on a Bruker D8 Advance diffractometer (Vantec position-sensitive detector, Ge-monochromatized Cu $K_{\alpha 1}$ radiation, Bragg-Brentano geometry, DIFFRACT plus software) in the $2\theta$ range 10-152$^\circ$ with a step size of 0.02$^\circ$ (counting time 15 s per step). The slit system was selected to ensure that the X-ray beam was completely within the sample for all $2\theta$ angles.

## 2.4 Second harmonic generation (SHG) measurements.

The material was characterized by SHG measurements in reflection geometry, using a pulsed Nd:YAG laser ($\lambda$=1.064 µm). The SHG signal $I_{2\omega}$ was measured from the polycrystalline sample relative to an $\alpha$-quartz standard at room temperature in the Q-switching mode with a repetition rate of 4 Hz. To make relevant comparisons microcrystalline powders of NISO, NSSO and $\alpha$-quartz standard were sieved into the same particle size range because the SHG efficiency depends strongly on particle size.

## 2.5 Magnetic and dielectric measurements

The magnetization experiments were performed in a Quantum Design MPMSXL 5 T SQUID magnetometer. The magnetization was recorded as a function of temperature using zero-field-cooled (ZFC) and field-cooled (FC) protocols. The magnetization was also recorded as a function of the magnetic field at different temperatures.

Dielectric properties of an NTO single crystal and of NSSO and NISO ceramic samples (0.3 mm thick disks) were measured using silver electrodes fired on two opposite sides of the material. Electrical contacts to the small faces of the samples were applied by firing a Pt-containing paste at 1170 K. The dielectric constant and loss tangent were measured using an R-5083 analyzer at different frequencies (100 kHz and 10-MHz) and dielectric spectroscopy measurements (Agilent 4284 A LCR meter, 1 V) were made in the temperature range from 300 to 1100 K, and frequencies from 100 Hz to 1 MHz.

**2.6 Thermogravimetric studies**

The thermogravimetric and calorimetric (TG-DTA/DSC) measurements were performed up to 1150 C under an oxygen atmosphere using a TG-DTA/DSC apparatus NETZSCH STA 409 PC/4/H.

**2.7 Neutron powder diffraction**

Because the neutron scattering lengths of Ni, Sc(In) and Sb are quite different, the chemical composition can be observed by neutron powder diffraction (NPD) with good precision ($b_{Ni}$ = 10.3 fm, $b_{Sc}$ = 12.29 fm, $b_{In}$ = 4.07 fm, $b_{Sb}$ = 5.57 fm). The neutron scattering length of oxygen ($b_O$ = 5.803 fm) is comparable to those of the heavy atoms and NPD provide accurate information on its position and stoichiometry. Several sets of NPD data were collected at the Institute Laue-Langevin (Grenoble, France) on NSSO and NISO samples and with different emphasis (high intensity or high resolution). High resolution data were collected on the powder diffractometer D2B (wavelength 1.5945 Å) with a step size of $0.05°$. The samples were inserted in a cylindrical vanadium container and a helium cryostat was used to collect data at several decisive temperatures in the temperature range 5-295 K. These data sets were used for the main crystallographic refinements. In order to follow the magnetic phase transitions with a much finer temperature step powder diffraction data were measured down to 1.7 K using the high intensity powder diffractometer D1B (wavelength 2.52 Å) equipped with a PSD covering the angular range 0.8 to 128.8º. D20 is a very high intensity 2-axis diffractometer equipped with a large microstrip detector (wavelength 2.422 Å). The scattering angle coverage of the detector was $10° < 2\theta < 150°$. Due to the extremely high neutron flux at D20, it was possible to resolve also very weak magnetic satellites.

Nuclear and magnetic refinements were performed by the Rietveld method using the FULLPROF software [36]. The diffraction peaks were described by a pseudo-Voigt profile function, with a Lorentzian contribution to the Gaussian peak shape. A peak asymmetry correction was made for angles below $35°$ ($2\theta$). Background intensities were estimated by interpolating between up to 40 selected points (low temperature NPD experimental data) or described by a polynomial with six coefficients. Because the studied structure is non-centrosymmetric the fractional coordinate z of one anion was fixed at 0.25 to define the origin along c-axis. An analysis of coordination polyhedra of cations was performed using the IVTON software [37]. The magnetic propagation vector was determined from the peak positions of the magnetic diffraction lines using the K-search software which is included in the FULLPROF refinement package [36]. Magnetic symmetry analysis was done using the program BASIREPS, also included in the FULLPROF suite [38]. The different allowed models for the magnetic structure were one by one tested against the measured data. Each structural model was refined to convergence, with the best result selected on the basis of agreement factors and stability of the refinement.

## 3. Results

### 3.1 Composition and grain size

It was found that phase formation of NSSO and NISO starts in the temperature range between 900 and 1000 C and single-phase samples have been obtained after annealing at 1250-1300C. According to the elemental analyses performed on 20 different crystallites, the metal compositions of NSSO and NISO are $Ni_{1.97(2)}Sc_{1.02(2)}Sb_{1.01(2)}$ and $Ni_{1.98(2)}In_{1.00(2)}Sb_{1.02(2)}$, if the sum of the cations is assumed to be 4. The oxygen content determined with iodometric titration was 5.98(3) and 5.99(3), for NSSO and NISO, respectively. All these values are very close to the expected ratios and permit to conclude that the sample compositions are the nominal ones ($Ni_2In(or Sc)SbO_6$). The microstructure of the obtained powders, observed by scanning electron microscopy, reveals uniform particle size distribution with an average grain size of 540 Å.

### 3.2 TG-DTA measurements

NTO, NISO and NSSO samples were subjected to TG analysis under flowing argon atmosphere at a heating rate 5 K min$^{-1}$. No weight loss was registered for the samples in temperature range 295-1130 K. Around 1130 K the TG signal for the NTO sample exhibited a weight loss peak. XRPD pattern of the decomposing mixture was taken after this run and it was found the set of peaks due to NTO and NiO. This is a good agreement with results reported in [39, 40] where a volatilization of the compound takes place at 1133 K. No such decompositions were observed in the NSSO and NISO case. However, as seen in the right panels of figure 1, DSC measurements revealed first order diffuse phase transitions around 1130 K and 1173 K for NSSO and NISO respectively. Unfortunately those temperatures exceed the measurement range of our current SHG and dielectric setups.

### 3.3 SHG measurements

Optical SHG measurements at room temperature for NISO, NSSO and NTO powdered samples gave distinct signals at $2\omega$ ($\lambda = 0.532$ nm) with intensity larger then that of quartz (from 1 to 4 of a QU (quartz unit)). Taking into account that all nickel oxides were intensively colored (thus making the output SHG intensity lower) one may conclude that characteristic SHG intensities for these substances are somewhat larger, amounting to order of 4-8 QU. This magnitude is quite sufficient to classify the crystal structures of the tested compounds as non-centrosymmetric [40] in accordance the crystal structure determination (polar *s.g. R3*). The established ferroelectric material $LiNbO_3$ with *R3c* symmetry and a room temperature spontaneous polarization, $P_s$, of 50 μC/cm$^2$ and an SHG signal of amplitude 220 QU. In the symmetry point groups 3 and 3m all non-vanishing third-order tensor elements represent so called vector parts of the optical nonlinearity susceptibility coefficient, $\delta^v$, which is directly related to the spontaneous polarization, $P_s$. Using the relations: $I_{2\omega} \sim (\delta^v)^2$ and $\delta^v \sim P_s$, it becomes possible to construct for $LiNbO_3$ and any other iso-symmetrical oxide X the simple relation: $P_s(X) = P_s(LiNbO_3) [I_{2\omega}(X)/ I_{2\omega}(LiNbO_3)]^{1/2}$ which for NISO and NSSO yields $P_s \sim$

$7\pm 2$ µC/cm$^2$. Though the SHG signal diminish with rising temperature it never drops to zero up to 1100 K in the case of NSSO and 1030 K for NISO (see left panels of figure 1), these ultimate temperatures are obviously lower than the expected phase transition into centro-symmetric states. Similar to the ferroelectric transformation at around 1300 K in LiNbO$_3$, these phase transitions may be of the second order also for NISO and NSSO.

### 3.4 Dielectric measurements

Dielectric properties of all sintered samples were measured at different frequencies (0.1 kHz<$f$<1 MHz) in the temperature range 300 <$T$<1100 K. The behaviour of the dielectric constant and loss factor for a single crystal of NTO is shown in figure 2. Sharp dielectric peaks are observed around 1000 K, at a frequency independent temperature. These peaks are assumed to correspond to a ferroelectric phase transition. The loss factor shows an increase in value up to the phase transition temperature followed by a slight decrease and then again an increase with increasing temperature. It should be noted that for ceramic NTO samples the temperature dependence of the dielectric constant depicts a typical relaxor behavior at lower temperatures, where a broad dielectric maximum shifts toward higher temperature with increasing frequency. The same type of relaxor behaviour (as for NTO ceramics) was registered in the dielectric permittivity curves for NISO and NSSO ceramic samples at temperatures lower than 1000 K. The observed relaxor behaviour may be related to a not large enough density of the available ceramics.

### 3.5 Magnetic measurements

The temperature-dependent magnetization curves of NSSO and NISO are shown in the upper panel of figure 3; the corresponding data for NTO is added for comparison. The shape of the magnetization curves suggests antiferromagnetic transitions for all compounds. Considering the maximum slope of M/H x T vs. T, we can estimate Neel temperatures, $T_N$, of 52 K for NTO, 60 K for NSSO, and 76 K for NISO, respectively. The high-temperature (T > 80 K) H/M data increases linearly with the temperature, indicating a Curie-Weiss behavior with Curie-Weiss temperatures θ ~ -49 K, -120 K, and -184 K respectively (see also Table 1). Hence, the magnitude of the antiferromagnetic interaction increases from NTO to NISO, akin to the increase of $T_N$. The frustration parameter f = - θ/$T_N$ is about 1 for NTO, while it is about 2 for NSSO and 2.4 for NISO, implying increased magnetic frustration in the substituted samples. The effective bohr magneton numbers $p_{eff}$ (see Table 1) obtained in the Curie-Weiss analyses are close to the usually observed value of 3.2 µ$_B$ for Ni$^{2+}$ in a 3d$^8$ configuration (considering S=1, L=0, $p_{eff}$=2.83 µ$_B$). As seen in the lower panels of figure 3, the magnetization is essentially linear with the applied magnetic field, as expected for an antiferromagnetic material. The weak increase of the susceptibility with increasing field below the transition temperature, resolved in the inset of figure 3, reflects the weak in-plane anisotropy of the incommensurate spiral structure of the magnetic moments (derived from the low temperature NPD measurements discussed below)

## 3.6 XRPD: Structural study

Crystallographic characterization of the samples was performed by X-ray powder diffraction analysis at room temperature which showed that the prepared samples of NISO and NSSO formed single phase powders. The room temperature XRPD patterns could be indexed on the basis of a trigonal unit cell with $a = 5.2168(1)$ Å, $c = 14.0167(3)$ Å for $Ni_2InSbO_6$ and $a = 5.1678(1)$ Å, $c = 14.0118(3)$ Å for $Ni_2ScSbO_6$. The data could be successfully refined by the Rietveld method using s.g. *R3*. As a starting model the crystal structure of corundum was used. All four *3a* sites are chemically equivalent, each forming 3 long bonds (2.1 A) and 3 short bonds (1.98 A) in a distorted octahedron. Any combination of the site occupancies within the overall 2Ni:1Sc(or In):1Sb composition is hence possible in principle. In the first stage of refinements a model with a random distribution of the Ni or Sc cations on available Wyckoff *3a* positions which are occupied by Ni and Te (Sb for NISO and NSSO). We used a set of 7 linear restrains where the fractional site occupancies were refined saving as a fixed value the full nominal limits for occupation and total content of cations on the 4 sites. It was found that Sc (or In) cation fully replace Ni in the 3a position of Ni(2) in the NTO structure. The other 3a sites Ni(1), Ni(3) and Sb were without anti-site disorder in a frame of standard deviation. The cation-oxygen bond lengths calculated from the refined lattice parameters and atomic coordinates are in close agreement with the literature data (ICSD data base). Furthermore, the corresponding bond valence sum calculations are consistent with the presence of $Ni^{2+}$, $In^{3+}$, $Sc^{3+}$, $Sb^{5+}$ and $O^{2-}$ ions. The final results from the Rietveld refinements with observed, calculated and difference plots for XRPD pattern of NTO, NISO and NSSO at 295 K are shown in figure 4.

## 3.7 NPD: Nuclear structure

The structural refinements of NISO and NSSO were carried out using both the high-resolution and the high-intensity neutron powder-diffraction data, recorded at room temperature. The data were fitted in the *R3* space group using the cell parameters of the XRPD analysis as a starting point. There is no significant mismatch between the observed intensity and the calculated profile. The structural parameters and the reliability factors of the refinement are summarized in Table 2 and the main interatomic distances are listed in Table 3.

The NSSO and NISO structures contain two crystallographically distinct Ni sites and the third Ni position (Ni(2)) of NTO is occupied by In or Sc. The Sb cations occupy the Te positions. The structures are related to that of corundum ($Al_2O_3$) with Ni, Sc (or In) and Sb replacing Al in a special ordered way. All metal cations occupy 3a (0 0 z) sites (with symmetry 3), while anions occupy the general position sites 9b. The anions in this structure are approximately hexagonally close-packed and within this framework the cations occupy 2/3 of the available octahedral holes. Each cation in this structure is surrounded by six oxygen ions forming a slightly distorted octahedron. From figure 5 it can be seen that all cations are stacked along the c-axis in a regular manner as the columns Ni(1) - Sc(or In)- Ni(3) - Sb. It is important to note that nonmagnetic Sc (or In) and Sb cations create holes in the magnetic Ni sublattices and prevent direct exchange interaction between the magnetic moments. The Sc(or In) cations are directly linked with the Ni(3) cations along the c-axis. It can be seen from figure 5 that there are 3 anions between two neighbouring planes which are shifted relative to each other by 1/3 of the a-parameter. Ni(3) is occupied above the centre of the hexagon formed by Ni(1) and Sc(or In) octahedra. In the hexagonal planes Ni(1) and Sc(or In) are connected through two anions forming a practically planar honeycomb lattice of edge-sharing octahedra. There are some opposite shifts of these cations with respect to the plane.

### 3.8 NPD: Magnetic structure

As mentioned earlier the neutron diffraction patterns of NSSO and NISO at room temperature are purely nuclear and can be refined in space group *R3* indicating the paramagnetic state at this temperature. The neutron diffraction patterns recorded on the D2B diffractometer on lowering the temperature (see figure 6) showed the appearance of additional magnetic peaks. A more detailed search for all possible magnetic satellites below $T_N$ was done using D1B and D20 and confirmed the existence of long-range magnetic ordering below about 60 K for NSSO and below 76 K for NISO, as indicated by the magnetic susceptibility measurements. Below $T_N$ similar sets of weak magnetic peaks occur for NSSO and NISO (see figure 7). Some of the new magnetic peaks seem to overlap with nuclear reflections, while others appear at positions extinct in the paramagnetic pattern, as expected from an antiferromagnetic ordering. Subtracting the high temperature paramagnetic scattering profile from the scattering profile of the magnetic phase gives a difference plot that corresponds to purely magnetic scattering. This difference plot is shown in figure 7b in the region $0.5°A^{-1} < |Q| < 2°A^{-1}$. Due to the thermal expansion of the sample the lattice parameters are slightly different at T = 1.7 and 80 K, and hence there is a small shift of the nuclear Bragg peaks with temperature creating a characteristic 'negative-positive' intensity profile at the positions of the nuclear Bragg peaks.

The propagation vector of the magnetic structure was determined using the indexing program "K-search" which is part of the FULLPROF suite [36]. The satellites cannot be indexed as simple multiples of the crystal unit cell, suggesting that the magnetic structure is incommensurate with the crystal lattice. Using the 10 most intense magnetic reflections visible in the difference pattern (figure 7b) the magnetic propagation vector was determined to **k** = (0, 0.036(1), 0) for NSSO and **k** = (0, 0.029(1), 0) for NISO. The values of the components $\mathbf{k_y}$ of the wave vector have been refined using the FULLPROF program and show a deviation from the commensurate value only for the b component. Due to the hexagonal symmetry it is not possible to differentiate between magnetic propagation vectors having a component in *a* or *b* direction; we can e.g. not exclude a wave vector k = (0.036, 0, 0) for NSSO. Magnetic symmetry analysis was used [37] to determine the allowed irreducible representations for the magnetic propagation vector **k** = (0, 0.036, 0) in spacegroup *R3*. The Fourier coefficients describing possible spin configurations can be written as linear combinations of irreducible representations of the wave vector group. For the two magnetic Ni-sites on Wyckoff position 3a in NSSO and NISO there exist only one allowed irreducible representation having three basis vectors in directions of the crystallographic unit cell parameters. Trying different models it turned out that a simple helical model with spin components perpendicular to the axis of the helix described the measured magnetic Bragg peak intensities very well. It was possible to constrain the moment values of the two independent Ni-sites to the same but antiferromagnetically coupled value without affecting the quality of the refinement. This led to a magnetic model with only two independent parameters, namely the value of the $\mathbf{k}_y$ component of the magnetic propagation vector and the value of the total magnetic moment on both Ni-sites. In the case of incommensurately modulated magnetic structures it is in principle not possible to differentiate between helical or sin wave modulated structures using the refinement of powder diffraction data. Physical arguments as e.g. the refined magnetic moment values can sometimes speak in favor of one or the other model. In our refinements the helical model gives a constant magnetic moment of about 1.9 $\mu_B$, a value which is close to the expected value for $Ni^{2+}$. Using a sin-wave model to fit the data leads to a sin-wave modulation of the magnetic moment between zero and a maximum value of about 2.7 $\mu_B$, a value clearly too big for $Ni^{2+}$. This indicates that the magnetic structures of NISO and NSSO are helix-like. It is important to underline that in hexagonal systems one does not know the direction of the magnetic component within the hexagonal basal plane. This means that we cannot obtain experimental evidence of a difference between a helix where the magnetic moments rotate within the *ac*-plane or a spiral (in fact a cycloidal spiral as the propagation vector is along b) where the moments rotate within the *bc*-plane. But we can say without any doubt that there is a moment component in c-direction excluding a spiral with magnetic moments just in the *ab*-plane. The refined moments of $\mu_{Ni}$ = 1.89(5) $\mu_B$ (for NISO) and 1.87(5) $\mu_B$ (for NSSO) at 1.7 K, are in good agreement with results in other NPD studies of Ni-based complex metal oxides [42]. The derived antiferromagnetic ordering is visualized in figure 9, where the Ni spins are constrained within the *ac*-plane and form a spiral arrangement that propagates along the *b*-direction.

## 4. Discussion and conclusions

It is found that the nuclear structure of NISO and NSSO retains *R3* symmetry down to 1.7 K. The evolution of the lattice constants with temperatures occurs without anomalies indicating structural transitions. Results from polyhedral analyses of the different cations in NTO, NSSO and NISO in the paramagnetic and the AFM phases are presented in Table 4. Compared to NTO, the cation-anion bond lengths in NISO and NSSO are increased for all four cation sublattices (Ni(1), Sc(In), Ni(3) and Sb) and also the polyhedral volume is increased. The magnetic cations are shifted away from the centroid of their coordination polyhedron, for NISO and NSSO this effect is accompanied by an increased polyhedral distortion compared to that of NTO. The non-magnetic cations have also moved away from the octahedral centers but these shifts are smaller than the magnetic cation shifts. It is found that the polyhedral volume of the Sb sites is smaller than those of the Ni and Sc(In) sites in agreement with the ionic radii of the cations. Results from bond-valence calculations [43,44] of the lattice data for NISO and NSSO are consistent with the presence of $Ni^{2+}$, $In^{3+}$, $Sc^{3+}$, $Sb^{5+}$ and $O^{2-}$ ions. There are asymmetric changes in the lattice parameters with doping where the a-parameter increases significantly more than the c-parameter. In all compounds, the value of hexagonal distortion *c/a* is smaller than for the ideal corundum structure (2.833): 2.695 for NTO, 2.687 for NISO and 2.711 for NSSO.

Our results show that the magnetic properties of NISO and NSSO are dependent of the nature of the nonmagnetic cations In and Sc. To explain this finding, we again consider the effects of In(Sc)/Sb substitutions on the crystal structure of NTO. The substitution expands the crystal lattice of NTO creating some additional structural distortions (see Table 4). The doping of nonmagnetic cations at the Ni sites decreases not only the number of exchange interaction pathways along the c-direction, but also inside of hexagonal planes. Nevertheless, the NISO and NSSO compounds are, as reported above, considerably more spin-frustrated than NTO. The low-temperature magnetic ordering should arise from several competing magnetic interactions. The outcome of this competition is likely to be modified because of the suppression of magnetic coupling by the deleting of one Ni cation. Following [23] the magnetic properties of NTO could be described by five different Ni-O-Ni spin superexchange parameters $J_1$-$J_5$. In the case of NISO and NSSO structures all FM contributions ( $J_1$ and $J_2$ paths ) should vanish upon doping, leaving only the strong AFM exchanges $J_3$-$J_5$, i.e. the doping stabilizes the AFM state and suppresses FM order in the samples. This modification may be as a reason for increasing of $T_N$ in the case of NISO and NSSO. Degradation of FM state after doping can be clearly predicted after the comparison of crystal structures of several compounds with corundum-related structure (hematite $Fe_2O_3$, NTO and NISO (NSSO)) which are depicted in figure 9. With increasing doping, the initial FM spin structure (hematite) transforms to AFM collinear order (NTO) and finally to the incommensurate AFM structure of NISO (or NSSO), as "magnetic planes" are removed from the structure.

The incommensurate nature of the propagation vector suggests that the spin interactions are frustrated along the b-direction. In [45] a mechanism for helical spin order is suggested which incorporates interplay between excess moments in non-ideal antiferromagnets and Dzyaloshinskii-Moriya anisotropy, which can occur because of a non-centrosymmetic structure [46]. Such a mechanism is possible in NISO and NSSO, but in addition, anisotropic exchange, dipolar interaction and crystal field anisotropy also play roles in determining the characteristics of helical magnetic structures [46,47].

NISO and NSSO adopt trigonal (*R3*) corundum-type structure at room temperature, and retain this symmetry down to 1.7 K.  A main feature of the structure is a special type of ordering where the cations are stacked along the c-axis as columns of Ni(1) - Sc(or In)- Ni(3)- Sb occupying 2/3 of the available octahedral holes. Magnetisation and NPD experiments evidence the existence of long range magnetic order in NISO and NSSO compounds at low temperature. The magnetic order is propagated with the incommensurate wave vectors **k** = (0, 0.036, 0) (for NSSO) and **k** = (0, 0.029, 0) (for NISO). In the NSSO case, the incommensurate component of the propagation vector leads to a spiral with a length of 28 unit cells (around 146 Å) and a turn angle of 12.6° between spins in neighbouring cells. No incommensurate modulation is present along the c axes and the repeat unit is equal to the lattice constant c. The appearance of an incommensurate antiferromagnetic spiral structure with a higher $T_N$ than that of NTO, is connected to a doping induced enhanced antiferromagnetic interaction and an accompanying spin frustration. It should be noted that the magnetic configuration could induce a (so-called Type II, [10]) polarization, if a finite magnetization component exists along the b-axis, in addition to the (Type I) polarization established at high-temperature [46].

## Acknowledgements

Financial support of this research from the Swedish Research Council (VR), Göran Gustafsson Foundation, the Swedish Foundation for International Cooperation in Research and Higher Education (STINT) and the Russian Foundation for Basic Research is gratefully acknowledged.

**Table 1**

| Compound | $\langle r_A \rangle$ | $r_B$ | $T_N$ (K) | $\theta_{CW}$ (K) | $p_{eff}$ ($m_B$) | Ferroelectric $T_c$ (K) |
|---|---|---|---|---|---|---|
| $Ni_3TeO_6$ | 0.69 | 0.56 | 52 | -49 | 3.08 | 1000 |
| $Ni_2ScSbO_6$ | 0.71 | 0.60 | 60 | -120 | 3.22 | > 1050 |
| $Ni_2InSbO_6$ | 0.73 | 0.60 | 76 | -184 | 3.34 | > 1050 |
| Compound | $\langle r_A \rangle$ | $r_B$ | $T_N$ (K) | $\theta_{CW}$ (K) | $p_{eff}$ ($m_B$) | Antiferro-electric $T_{AFE}$ (K) |
| $Mn_2ScSbO_6$ | 0.80 | 0.60 | - | -76 | 5.79 | 470 |
| $Mn_2InSbO_6$ | 0.82 | 0.60 | - | -64 | 5.87 | 485 |
| $Mn_3TeO_6$ | 0.83 | 0.56 | 24 | -120 | 5.93 | 510 |

**Table 2** Summary of the results of the structural refinements of the $Ni_3TeO_6$, $Ni_2InSbO_6$ and $Ni_2ScSbO_6$ samples using XRPD and NPD data (*s.g. **R3**, z=3*)

| Data | XRPD | | | | NPD | | | |
|---|---|---|---|---|---|---|---|---|
| Sample | $Ni_3TeO_6$ Ref. [35] | $Ni_3TeO_6$ | $Ni_2InSbO_6$ | $Ni_2ScSbO_6$ | $Ni_2InSbO_6$ | | $Ni_2ScSbO_6$ | |
| T,K | 292 | 295 | 295 | 295 | 295 | 1.7 | 295 | 1.7 |
| a[Å] | 5.1087(8) | 5.1058(3) | 5.2168(3) | 5.1678(3) | 5.2158(3) | 5.2101(3) | 5.1668(3) | 5.1614(3) |
| c[Å] | 13.767(1) | 13.7621(5) | 14.0166(4) | 14.0117(5) | 14.0139(5) | 14.0065(5) | 14.0080(5) | 14.0025(6) |
| Ni(1) (0,0,z) | | | | | | | | |
| z | 0.0127(8) | 0.0200(6) | -0.0139(9) | -0.0275(8) | -0.0196(6) | -0.0201(6) | -0.0211(6) | -0.0177(8) |
| $B[Å]^2$ | 0.0064 | 0.36(4) | 0.68(4) | 0.62(4) | 0.43(3) | 0.35(3) | 0.47(3) | 0.39(3) |
| Ni(2) (or Sc,In) (0, 0, z) | | | | | | | | |
| z | 0.2144(5) | 0.2255(5) | 0.1938(9) | 0.1773(9) | 0.1827(6) | 0.1864(6) | 0.1849(6) | 0.1874(6) |
| $B[Å]^2$ | 0.0093 | 0.67(4) | 0.54(4) | 0.67(4) | 0.53(3) | 0.42(3) | 0.44(3) | 0.57(3) |
| Ni(3) (0, 0, z) | | | | | | | | |
| z | 0.5066(3) | 0.5087(7) | 0.4798(8) | 0.4671(8) | 0.4729(5) | 0.4753(5) | 0.4768(5) | 0.4783(5) |
| $B[Å]^2$ | 0.0061 | 0.45(1) | 0.55(4) | 0.51(4) | 0.67(3) | 0.54(3) | 0.64(3) | 0.52(3) |
| Te or Sb (0, 0, z) | | | | | | | | |
| z | 0.7085(5) | 0.7125(5) | 0.6845(8) | 0.6698(8) | 0.6799(5) | 0.6787(5) | 0.6752(5) | 0.6770(5) |
| $B[Å]^2$ | 0.0041 | 0.38(1) | 0.49(4) | 0.37(4) | 0.63(6) | 0.51(2) | 0.47(3) | 0.38(2) |
| O(1) | | | | | | | | |
| x | 0.3263(3) | 0.3163(6) | 0.3010(8) | 0.3052(9) | 0.3204(7) | 0.3238(8) | 0.3222(8) | 0.3164(8) |
| y | 0.0250(4) | 0.0017(5) | 0.0089(9) | 0.0081(8) | 0.0132(8) | 0.0152(7) | 0.0194(7) | 0.0147(7) |
| z | 0.1120(4) | 0.0942(7) | 0.0814(8) | 0.0781(9) | 0.0764(8) | 0.0770(6) | 0.0739(9) | 0.0749(6) |
| $B[Å]^2$ | 0.0064 | 1.26(3) | 1.18(5) | 1.09(8) | 0.87(5) | 0.81(4) | 1.31(3) | 1.08(4) |
| O(2) | | | | | | | | |
| x | 0.3301(3) | 0.3152(5) | 0.3711(9) | 0.3458(9) | 0.3436(8) | 0.3471(9) | 0.3329(7) | 0.3324(7) |

| *y* | 0.3702(3) | 0.3412(6) | 0.3707(8) | 0.3697(8) | 0.3817(9) | 0.3861(8) | 0.3726(8) | 0.3692(8) |
|---|---|---|---|---|---|---|---|---|
| *z* | 0.2845(3) | 0.25 | 0.25 | 0.25 | 0.25 | 0.25 | 0.25 | 0.25 |
| *B[Å]$^2$* | 0.0041 | 1.16(3) | 1.29(4) | 1.32(6) | 1.12(6) | 0.96(6) | 1.13(6) | 0.96(5) |
| *kz* | | | | | | -0.029(2) | | -0.036(2) |
| *μ, μ$_B$* | | | | | | 1.89(2) | | 1.86(2) |
| R$_p$,% | | 6.11 | 6.33 | 4.81 | 5.10 | 4.34 | 5.43 | 5.31 |
| R$_{wp}$,% | | 7.54 | 7.66 | 7.06 | 7.03 | 6.18 | 7.39 | 7.18 |
| R$_B$,% | 2.4 | 6.18 | 5.98 | 5.24 | 5.61 | 5.43 | 5.39 | 5.11 |
| R$_{mag}$,% | | - | - | - | - | 6.8 | - | 11.3 |
| χ$^2$ | | 1.93 | 2.13 | 2.23 | 2.19 | 2.26 | 2.15 | 2.32 |

**Table 3.** Selected bond lengths [Å] of Ni$_3$TeO$_6$, Ni$_2$ScSbO$_6$ and Ni$_2$InSbO$_6$ samples.

| Sample | Anion | Ni$_2$InSbO$_6$ | | Ni$_2$ScSbO$_6$ | | Ni$_3$TeO$_6$ |
|---|---|---|---|---|---|---|
| T,K | Oxygen | NPD 295 | NPD 1.7 | NPD 295 | NPD 1.7 | XRPD 295 |
| Ni(1) | O(2) | 2.031(6) (x3) | 2.022(6) (x3) | 2.043(5) (x3) | 2.031(7) (x3) | 2.008(5) (x3) |
|  | O(1) | 2.120(6) (x3) | 2.137(6) (x3) | 2.093(5) (x3) | 2.057(6) (x3) | 2.110(5) (x3) |
| Ni(2) (Sc, In) | O(2) | 2.121(6) (x3) | 2.115(6) (x3) | 2.046(5) (x3) | 2.018(7) (x3) | 2.043(5) (x3) |
|  | O(1) | 2.264(6) (x3) | 2.251(6) (x3) | 2.254(5) (x3) | 2.243(6) (x3) | 2.135(5) (x3) |
| Ni(3) | O(1) | 2.054(6) (x3) | 2.044(6) (x3) | 2.093(5) (x3) | 2.087(5) (x3) | 2.032(5) (x3) |
|  | O(1) | 2.165(6) (x3) | 2.130(5) (x3) | 2.131(5) (x3) | 2.124(6) (x3) | 2.149(5) (x3) |
| Te (Sb) | O(1) | 1.962(6) (x3) | 1.951(5) (x3) | 1.954(5) (x3) | 1.937(7) (x3) | 1.923(5) (x3) |
|  | O(2) | 2.031(6) (x3) | 2.008(5) (x3) | 2.017(5) (x3) | 1.990(6) (x3) | 1.956(5) (x3) |
| Ni(1)-Ni(3), Å | | 3.75(1) | 3.72(1) | 3.77(1) | 3.74(1) | 3.76(1) |
| Ni(1)-O(1)-Ni(3), deg. | | 128.3(6) | 128.1(6) | 128.0(6) | 127.6(6) | 131.4(6) |
| Ni(1)-O(2)-Ni(3), deg. | | 135.8(6) | 134.9(6) | 132.8(6) | 132.0(4) | 125.7(4) |

**Table 4** Polyhedral analysis of $Ni_3TeO_6$ (NTO), $Ni_2InSbO_6$ (NISO) and $Ni_2ScSbO_6$ (NSSO) at different temperatures (cn - coordination number, x – shift from centroid, $\xi$- average bond distance with a standard deviation, V- polyhedral volume, $\omega$ - polyhedral volume distortion).

| Sample | T,K | Cation | cn | x(Å) | $\xi$ (Å) | V(Å$^3$) | $\omega$ | Valence |
|---|---|---|---|---|---|---|---|---|
| NTO | 295 | | | 0.166 | 2.059+/-0.056 | 11.20(1) | 0.025 | 2.02 |
| | 295 | | | 0.172 | 2.063+/-0.035 | 11.30(1) | 0.027 | 1.99 |
| NSSO | 1.7 | Ni(1) | 6 | 0.176 | 2.050+/-0.008 | 11.18(1) | 0.024 | 1.98 |
| NISO | 295 | | | 0.144 | 2.080+/-0.062 | 11.59(1) | 0.022 | 1.96 |
| | 1.7 | | | 0.131 | 2.074+/-0.049 | 11.53(1) | 0.021 | 2.14 |
| NTO | 295 | Ni(2) | | 0.137 | 2.089+/-0.050 | 11.92(1) | 0.010 | 1.89 |
| NSSO | 295 | | | 0.149 | 2.144+/-0.108 | 12.68(1) | 0.012 | 2.95 |
| | 1.7 | Sc | 6 | 0.154 | 2.131+/-0.123 | 12.37(1) | 0.013 | 2.94 |
| NISO | 295 | | | 0.301 | 2.183+/-0.075 | 13.34(1) | 0.021 | 2.96 |
| | 1.7 | In | | 0.198 | 2.167+/-0.051 | 13.17(1) | 0.018 | 2.89 |
| NTO | 295 | | | 0.242 | 2.090+/-0.064 | 11.58(1) | 0.031 | 1.88 |
| NSSO | 295 | | | 0.228 | 2.111+/-0.020 | 12.13(1) | 0.028 | 1.89 |
| | 1.7 | Ni(3) | 6 | 0.212 | 2.106+/-0.020 | 12.08(1) | 0.024 | 1.96 |
| NISO | 295 | | | 0.226 | 2.104+/-0.066 | 11.86(1) | 0.027 | 1.89 |
| | 1.7 | | | 0.236 | 2.092+/-0.042 | 11.71(1) | 0.029 | 1.88 |
| NTO | 295 | Te | | 0.112 | 1.940+/-0.018 | 9.61(1) | 0.009 | 5.93 |
| NSSO | 295 | | | 0.147 | 1.986+/-0.030 | 10.23(1) | 0.015 | 4.96 |
| | 1.7 | Sb | 6 | 0.137 | 1.964+/-0.030 | 9.91(1) | 0.013 | 5.05 |
| NISO | 295 | | | 0.166 | 1.992+/-0.042 | 10.25(1) | 0.018 | 4.93 |
| | 1.7 | Sb | | 0.163 | 1.979+/-0.031 | 10.08(1) | 0.018 | 4.93 |

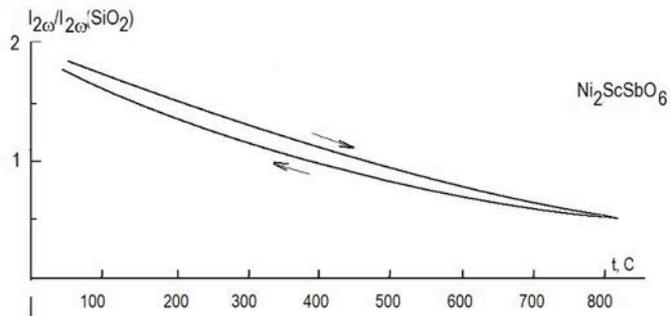
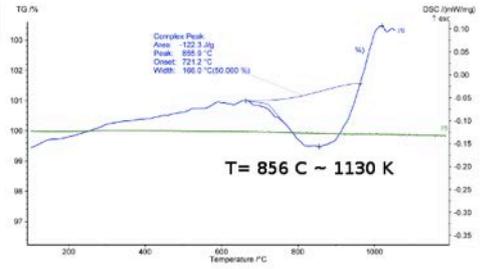
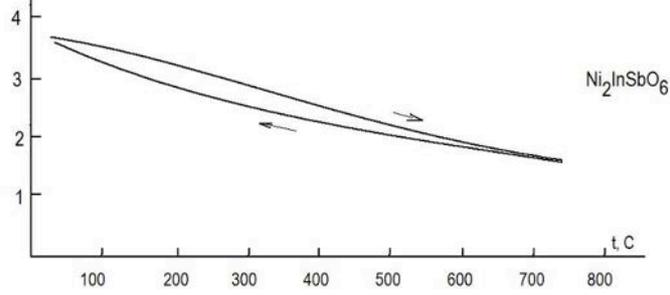
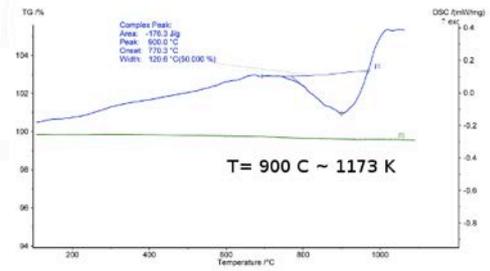

**Figure 1**

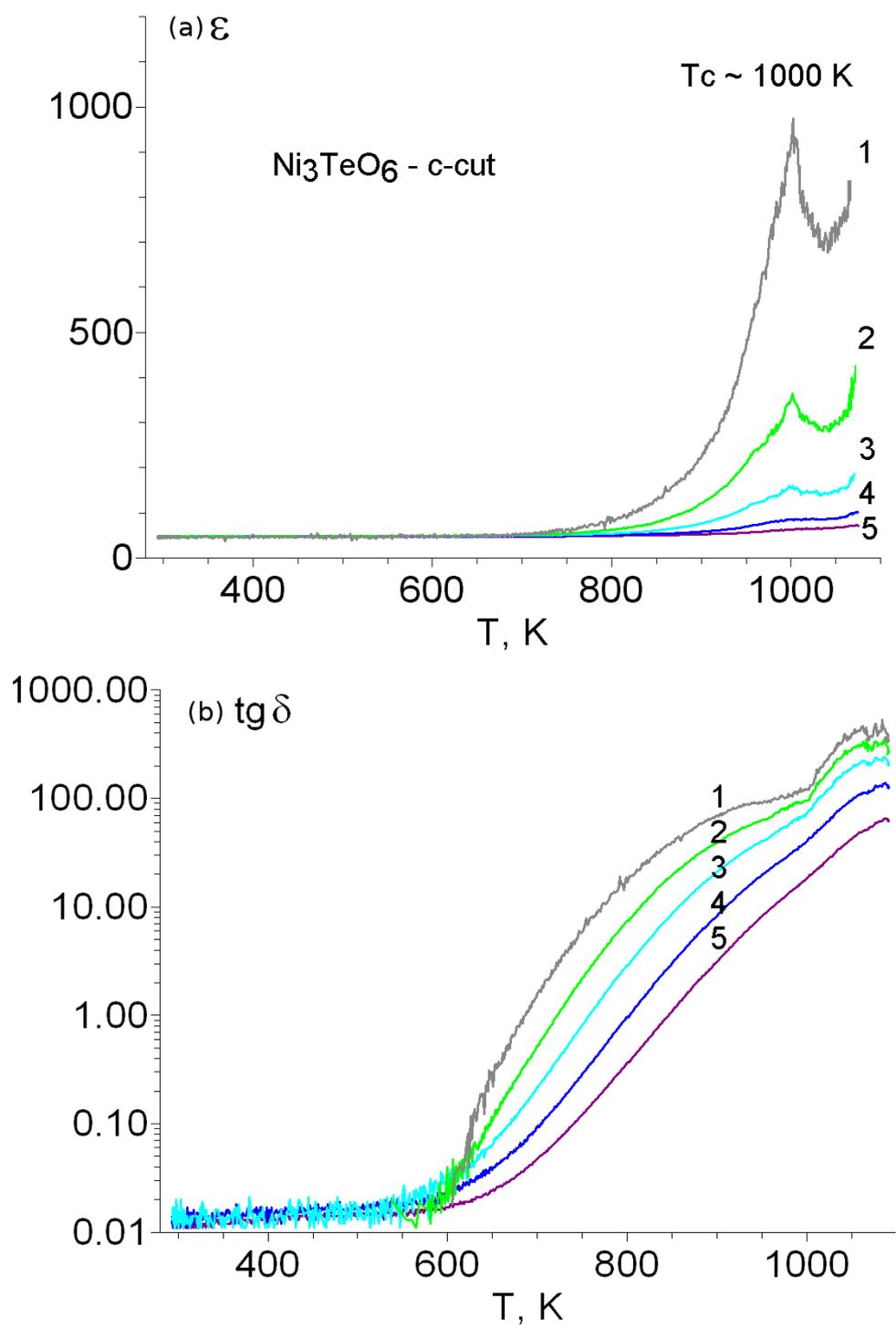

**Figure 2**

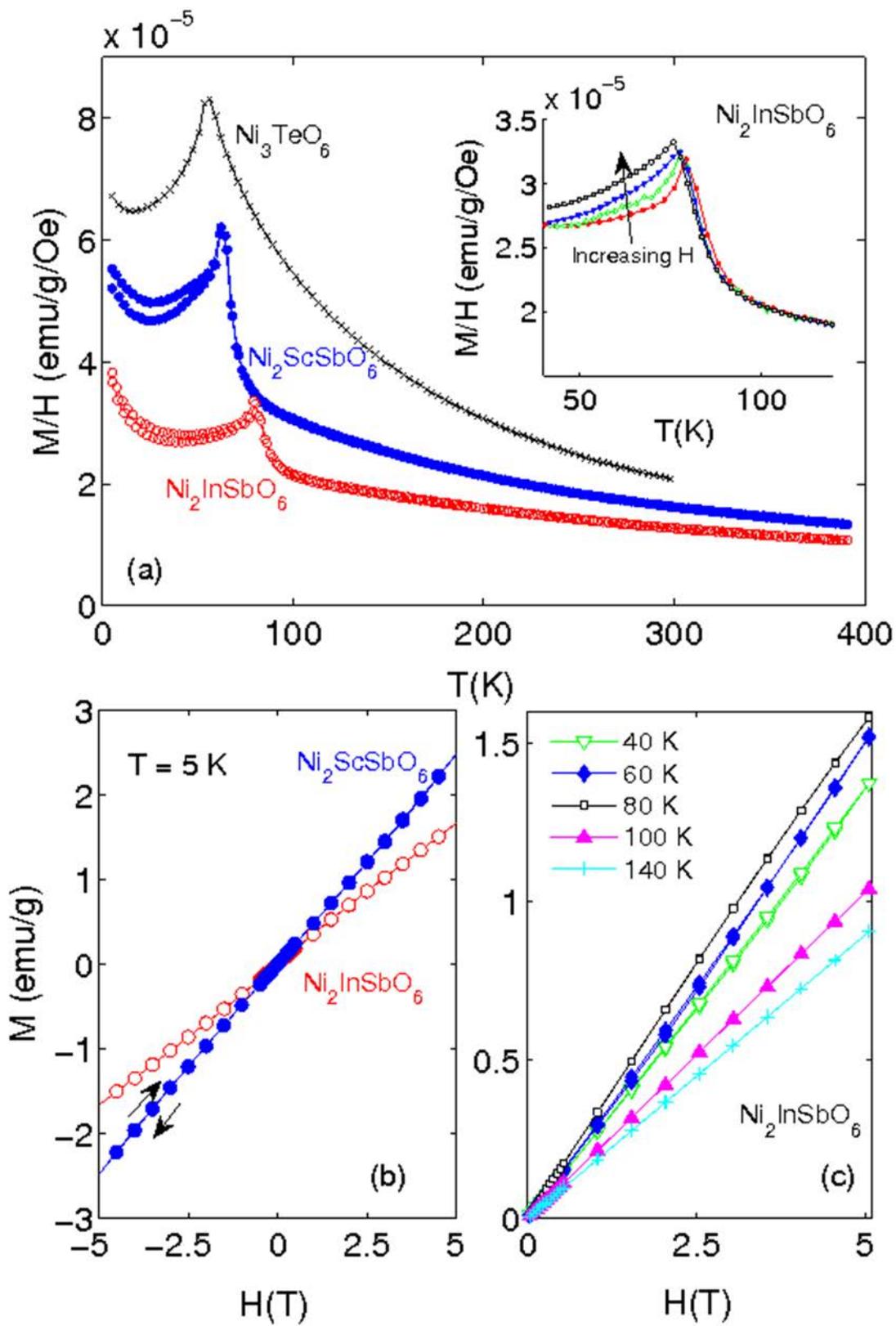

**Figure 3**

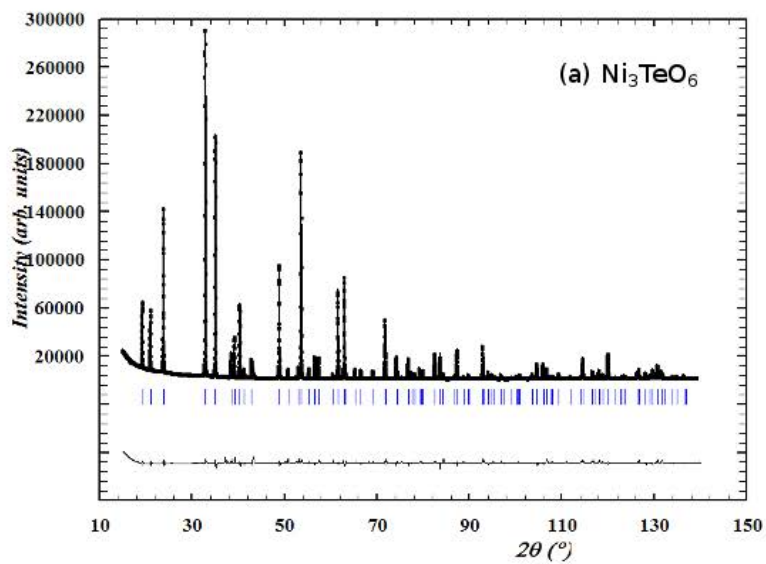

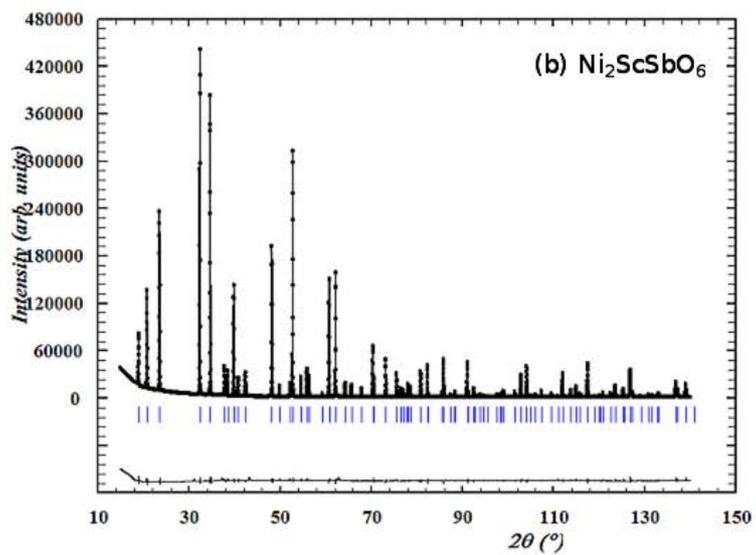

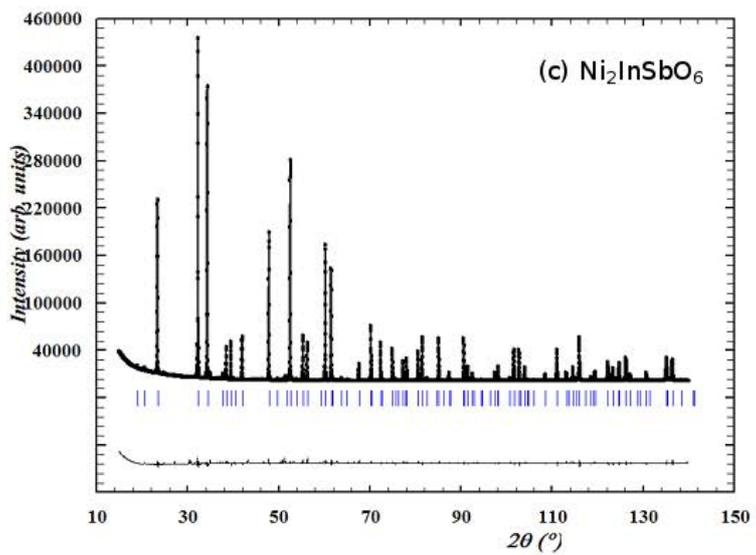

**Figure 4**

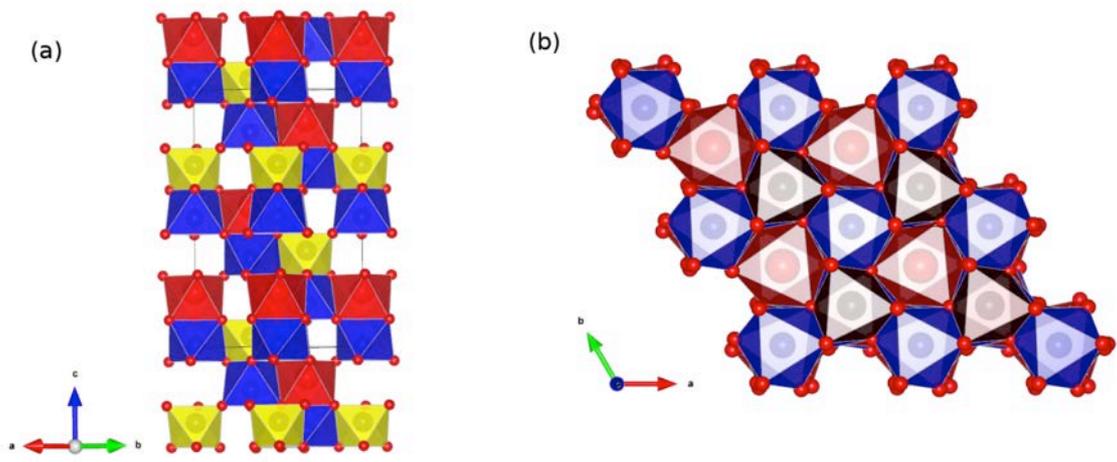

**Figure 5**

Ni2ScSbO6 D20 ILL

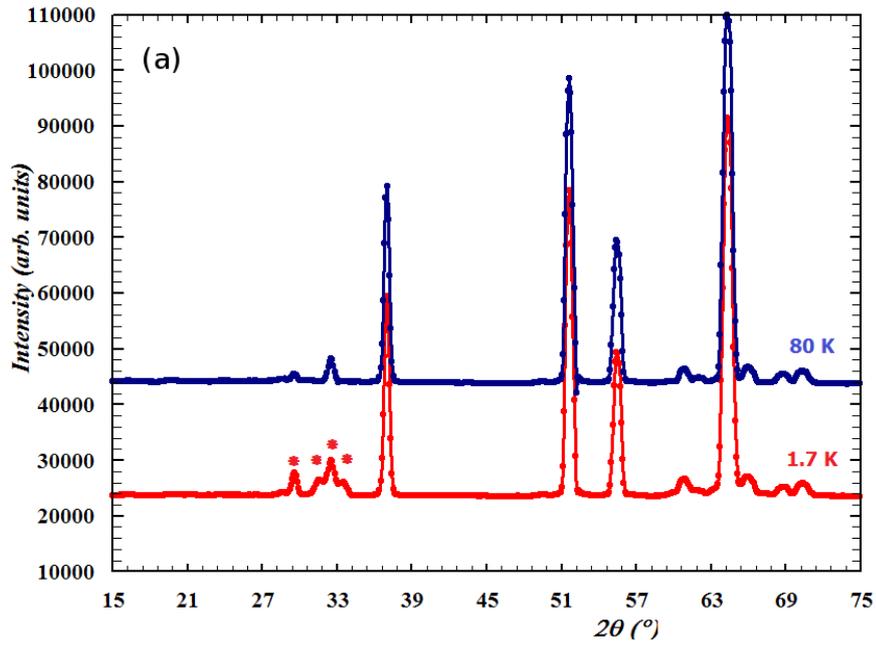

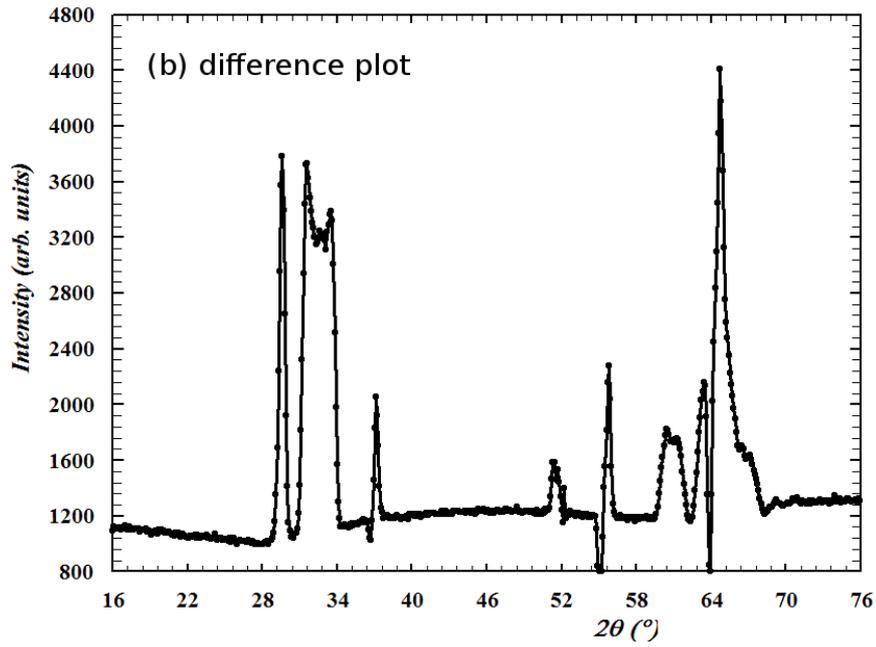

**Figure 6**

**Ni2ScSbO6**

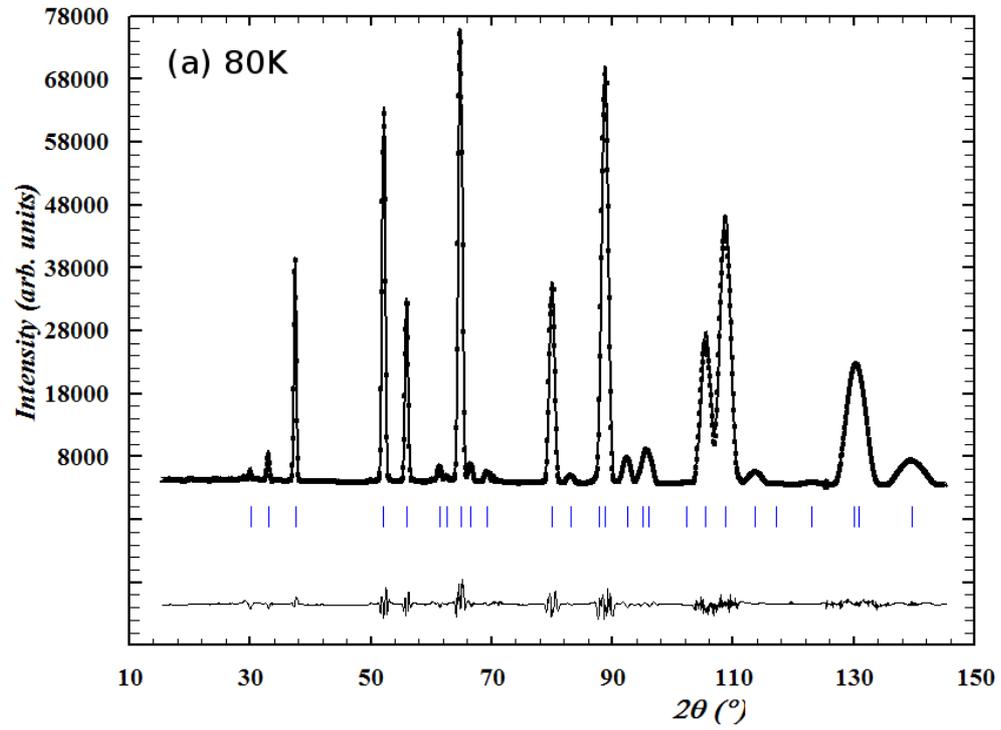

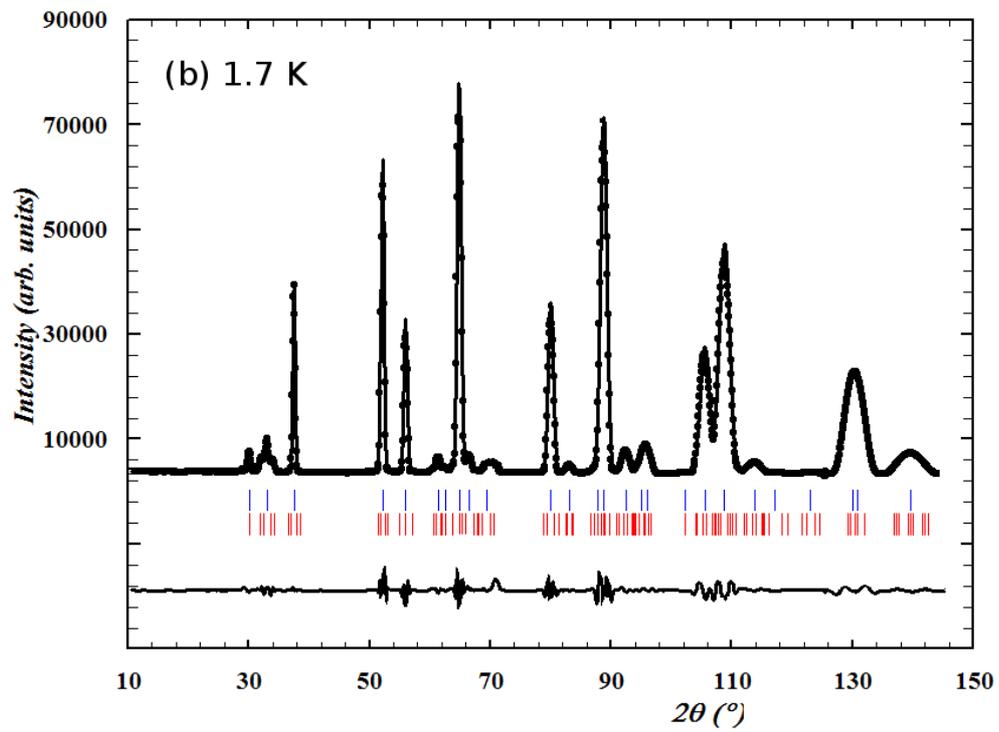

Figure 7

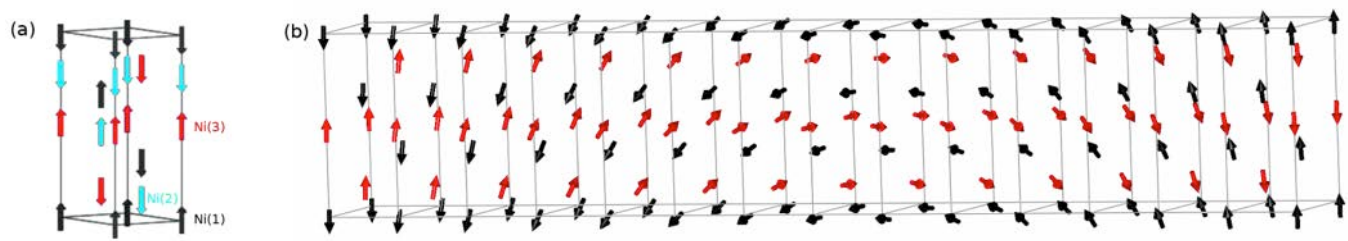

**Figure 8**

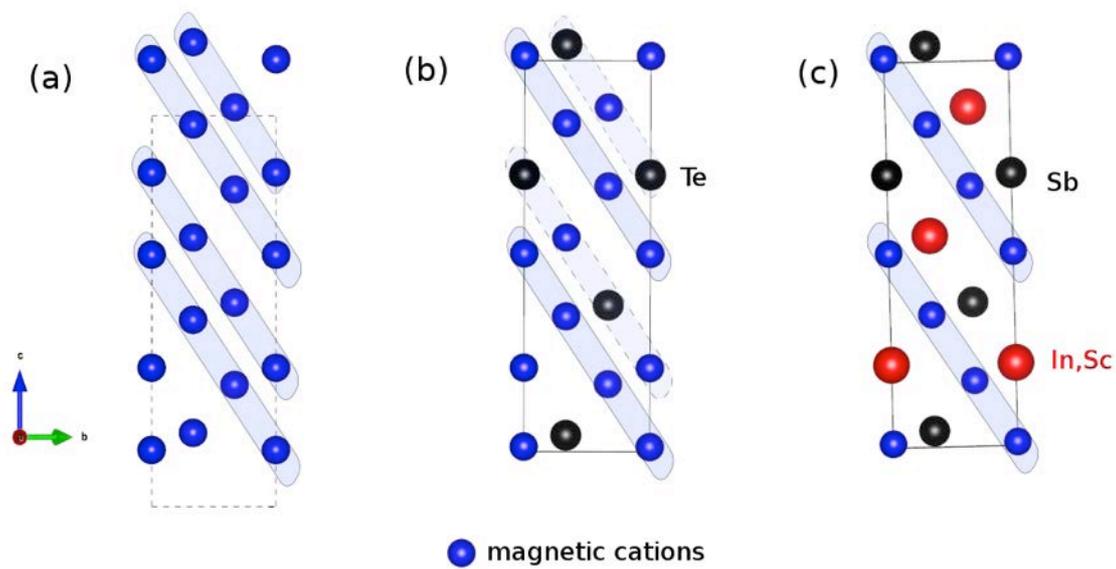

**Figure 9**

**Figure captions**

**Figure 1** Temperature dependence of (left) the second harmonic generation signal and (right) DSC data for NISO and NSSO samples.

**Figure 2** Temperature dependence of (a) dielectric constant $\varepsilon$ and (b) dielectric loss tgδ for an NTO single crystal at different frequencies: 100 Hz (1), 1 kHz (2), 10 kHz (3), 100 kHz (4), 1 MHz (5).

**Figure 3** (a): Temperature dependence of the ZFC/FC susceptibility, M/H, for $Ni_2ScSbO_6$ and $Ni_2InSbO_6$ (recorded in H = 20 Oe) and $Ni_3TeO_6$ (H=1000 Oe, only FC curve is shown). The inset shows the temperature dependence of the field-cooled M/H for $Ni_2InSbO_6$, recorded in 20 Oe, 0.5 T, 1.5 T and 5 T respectively. (b) and (c): Hysteresis curves at (b) T = 5 K for both $Ni_2ScSbO_6$ and $Ni_2InSbO_6$ and at (c) different temperatures for $Ni_2InSbO_6$. The arrows indicate the field-sweep directions; only positive magnetic field values are shown in (c).

**Figure 4** XRPD patterns for (a) NTO, (b) NSSO and (c) NISO at room temperature.

**Figure 5** (a) Polyhedral representation of the NISO (or NSSO) crystal structure; (b) honeycomb-like (ab) planes with hexagonal structure and connection between cation octahedra.

**Figure 6** NPD patterns of NSSO measured at (a) 80 K and 1.7K and (b) the associated difference plot.

**Figure 7** The observed, calculated, and difference plots for the fit to NPD patterns of NSSO after the Rietveld refinement of the nuclear and magnetic structure at different temperatures: (a) 80K and (b) 1.7 K.

**Figure 8** A sketch of the (a) collinear structure of $Ni_3TeO_6$ (adapted from Ref. [21]), and (b) incommensurate helical magnetic structure of NISO (or NSSO) at T = 1.7 K. Diamagnetic ions are omitted. The propagation vector **k** is along the b-axis.

**Figure 9** Different types of cation order in $Fe_2O_3$, $Ni_3TeO_6$ and $Ni_2Sc$(or In)$SbO_6$ with corundum structure.